\documentstyle[11pt]{article}

\def\kon#1#2{\vbox{\halign{##&&##\cr\lower4pt
\hbox{$\scriptscriptstyle\vert$}\hrulefill &\hrulefill\lower4pt
\hbox{$\scriptscriptstyle\vert$}\cr $#1$&$#2$\cr}}}

\def\fii{\varphi}
\def\al{\alpha}

\def\ro{\varrho}

\def\d{\partial}
\def\=d{\,{\buildrel\rm def\over =}\,}

\def\sqr#1#2{{\vcenter{\vbox{\hrule height.#2pt\hbox{\vrule width.
#2pt height#1pt \kern#1pt \vrule width.#2pt}\hrule height.#2pt}}}}

\def\te{\vartheta}
\def\B{\Bigl}
\def\el{{\rm el}}

\begin{document}

\title{Nonstandard cosmology }
\author{G\"unter Scharf
\footnote{e-mail: scharf@physik.uzh.ch}
\\ Institut f\"ur Theoretische Physik, 
\\ Universit\"at Z\"urich, 
\\ Winterthurerstr. 190 , CH-8057 Z\"urich, Switzerland}

\date{}

\maketitle\vskip 3cm

\begin{abstract} 
Considering radial geodesics in the Robertson-Walker metric leads us to abandon the co-moving coordinates. Instead we work in the cosmic rest frame. Since then the matter is in motion, the solution of Einstein's equations is more complicated. We calculate the first correction to standard cosmology which has an off-diagonal term $bdt\,dr$ in the metric. It describes the late universe. We then solve Maxwell's equations in the new metric and discuss redshift and luminosities. We obtain the correct age of the universe $T=$ 14 Gyr$=1/H_0$, without assuming a cosmological constant,
\vskip 1cm
{\bf Keyword: Cosmology }

\end{abstract}

\newpage

\section{Introduction}

In standard general relativity it has become customary to introduce hypothetical new sources of gravity if there is a problem to understand observations. So one postulates dark matter to understand the large circular velocities of outer stars and gas in galaxies. In cosmology one assumes dark energy to understand the apparent accelerated expansion of the universe and one introduces one or more inflaton fields to solve the horizon problem. Gradually there is an inflation of sources of gravity.

In nonstandard general relativity we do not introduce hypothetical sources. Instead we investigate more general solutions of Einstein's equations [1] [2] [3]. These solutions do not have a geometric interpretation, they describe the gravitational fields which are responsible for the observed phenomena. Geometry on the other hand is a convention. Taking over this non-geometric view to cosmology we ascribe the astronomical observations to an expanding cosmic gravitational field which governs the motion of stars, galaxies and radiation.

The paper is organized as follows. In the next section we study radial geodesics in the Robertson-Walker metric. This will lead us to abandon the co-moving coordinates and to work in the cosmic rest frame instead. But then in contrast to standard cosmology, the matter is in radial motion which must be taken into account in the energy-momentum tensor. The radial velocity defines a new universal constant $K$ which is characteristic for our universe. In Sect.3 we solve Einstein's equations for a matter dominated universe representing the matter by a perfect fluid. We calculate the first correction to the Robertson-Walker metric. To make contact with the observations we investigate in Sect.4 Maxwell's equation in the new metric. In this way we are able to identify the cosmological redshift and the luminosities. In the last section we use our results to discuss the redshift luminosity-distance relation. We get the correct age of the universe $T=14 $Gyr$=1/H_0$, without assuming a cosmological constant. The new constant $K$ determines the deceleration parameter $q_0$. A negative $q_0$ which corresponds to accelerated expansion is well possible without assuming dark energy.

\section{Radial geodesics}

We choose spherical coordinates $t,r,\te,\phi$ which are assumed to be dimensionless, that means the measured values have been divided by suitable units. The light velocity is not put equal to 1 because it will explicitly appear later on (Sect.4). We start from the Robertson-Walker metric with flat spatial part
$$ds^2=dt^2-a^2(t)[dr^2+r^2(d\te^2+\sin^2\te d\phi^2)].\eqno(2.1)$$
The components of the metric tensor are
$$g_{00}=1,\quad g_{ij}=-a^2\gamma_{ij}$$
$$\gamma_{ij}={\rm diag}(1,r^2,r^2\sin^2\te)\eqno(2.2)$$
$$g^{00}=1,\quad g^{ij}=-{1\over a^2\gamma_{ij}}.\eqno(2.3)$$
The corresponding non-vanishing Christoffel symbols are given by
$$\Gamma^0_{11}=-a\dot a\eqno(2.4)$$
$$\Gamma^0_{22}=-r^2a\dot a,\quad \Gamma^0_{33}=-a\dot ar^2\sin^2\te,\eqno(2.5)$$
$$\Gamma^1_{10}={\dot a\over a}\eqno(2.6)$$
$$\Gamma^1_{22}=-r,\quad \Gamma^1_{33}=-r\sin^2\te,\eqno(2.7)$$
$$\Gamma^2_{20}={\dot a\over a},\quad\Gamma^2_{21}={1\over r},\eqno(2.8)$$
$$\Gamma^2_{33}=-\sin\te\cos\te,\eqno(2.9)$$
$$\Gamma^3_{03}={\dot a\over a},\quad\Gamma^3_{31}={1\over r},\eqno(2.10)$$
$$\Gamma^3_{23}={\cos\te\over\sin\te}.\eqno(2.11)$$
The dot always means time derivative. 

In nonstandard general relativity we do not interpret this metric. Instead we consider the geodesic equation
$${d^2x^\mu\over d\tau^2}+\Gamma^\mu_{\al\beta}{dx^\al\over d\tau}
{d x^\beta\over d\tau}=0.\eqno(2.12)$$
It describes the motion of test bodies which is certainly observable. We are interested in the radial geodesics $\te={\rm const.}, \phi={\rm const.}$.  Then the geodesic equations are simply given by 
$${d^2t\over d\tau^2}+\Gamma^0_{00}\B({dt\over d\tau}\B)^2+2\Gamma^0_{01}{dt\over d\tau}{dr\over d\tau}+
\Gamma^0_{11}\B({dr\over d\tau}\B)^2=0\eqno(2.13)$$
$${d^2r\over d\tau^2}+\Gamma^1_{00}\B({dt\over d\tau}\B)^2+2\Gamma^1_{01}{dt\over d\tau}{dr\over d \tau}+
\Gamma^1_{11}\B({dr\over d\tau}\B)^2=0\eqno(2.14)$$
As a first step we eliminate the affine parameter $\tau$. From
$${dr\over dt}={dr\over d\tau}{d\tau\over dt}\eqno(2.15)$$
we find
$${d^2r\over dt^2}=\B[{d^2r\over d\tau^2}{d\tau\over dt}+{dr\over d\tau}{d\over d\tau}\B({dt\over d\tau}\B)^{-1}\B]
{d\tau\over dt}=$$
$$={d^2r\over d\tau^2}\B({d\tau\over dt}\B)^2-{dr\over d\tau}\B({dt\over d\tau}\B)^{-2}{d^2t\over d\tau^2}{d\tau\over dt}=$$
$$={d^2r\over d\tau^2}\B({d\tau\over dt}\B)^2-\B({dt\over d\tau}\B)^{-2}{d^2t\over d\tau^2}{dr\over dt}.\eqno(2.16)$$
Substituting the geodesic equations (2.13) (2.14) inhere we get
$${d^2r\over dt^2}+\B(2\Gamma^1_{01}-\Gamma^0_{00}\B){dr\over dt}+\B(\Gamma^1_{11}-2\Gamma^0_{01}\B)\B({dr\over dt}\B)^2-\Gamma^0_{11}\B({dr\over dt}\B)^3+\Gamma^1_{00}=0.\eqno(2.17)$$
This is now a first order differential equation for the radial velocity
$$v_r(t)={dr\over dt}.\eqno(2.18)$$
As a consequence of the homogeneity assumption in the Robertson-Walker metric $v_r(t)$ does not depend on the radial coordinate $r$.

For the simple Robertson-Walker metric (2.1) the radial geodesic equation reads
$${dv_r\over dt}+2{\dot a\over a}v_r-a\dot av_r^3=0.\eqno(2.19)$$
This equation has the trivial solution $v_r=0$, which means $r=$const. This leads to the interpretation of $r,\te,\phi$ as co-moving coordinates and to standard cosmology. {\it We cannot believe that the Creator of the Universe has chosen this trivial solution;} in other less theological words: we consider the co-moving coordinates as unphysical. Therefore we look for non-trivial solutions of  equation (2.19).
Since this is an equation of Bernoulli type it can be easily integrated by the substitution
$$v_r(t)={u(t)\over a^2(t)}.$$
Then we have
$$\dot u={\dot a\over a^3}u^3\eqno(2.20)$$
with the solution
$$u={a\over\sqrt{K^2a^2+1}}\eqno(2.21)$$
so that
$$v_r(t)={1\over a\sqrt{K^2a^2+1}},\eqno(2.22)$$
where $K^2$ is a constant of integration. It must be positive to prevent $v_r$ to become imaginary for big $a(t)$. For $K=0$ we have the null geodesics which describe light rays. Since all massive test bodies fall with equal speed in the gravitational field, $K$ must be a universal constant in the true sense of the word. This constant $K$ which is characteristic for our universe has direct observational consequences: we shall see at the end of the paper that it determines the deceleration parameter.

The universe in nonstandard cosmology looks now as follows. With the radial velocity $v_r(t)$ (2.22) our coordinates are not co-moving but fixed, defined by the astronomers on earth. More precisely, one corrects for the motion of the earth so that the coordinate system is at rest with respect to the cosmic background radiation. This is the cosmic rest frame. The fact that we observe a rather small velocity of the solar system in this frame shows that the universe is very old, so that $v_r(t)$ (2.22)
is small for $t=T$ because $Ka(T)$ is big $\gg 1$ where $T$ is the age of the universe. The redshifted galaxies which we observe at the sky show that the radial velocity was indeed bigger in earlier times. Of course the radial motion of all matter must be taken into account in the energy-momentum tensor
$T^{\mu\nu}$ which is the source of the gravitational field. Consequently, instead of the static tensor used in standard cosmology we represent the matter by the energy-momentum tensor of a perfect fluid in a gravitational field:
$$T^{\mu\nu}=-pg^{\mu\nu}+(p+\ro)u^\mu u^\nu,\eqno(2.23)$$
the minus sign is due to our choice $(+,-,-,-)$ of the metric. Here $u^\mu$ is the 4-velocity
$$u^\mu=(u^0,u^0V^j)\eqno(2.24)$$
which is normalized
$$g_{\mu\nu}u^\mu u^\nu=1.\eqno(2.25)$$
The 3-velocity $V^j$ has $V^1=v_r$ given by (2.22) and $V^2=0=V^3$. Then (2.24) and (2.25) imply
$$u^0=\sqrt{1+{1\over K^2a^2}}\eqno(2.26)$$
and
$$u^1=u^0v_r={1\over Ka^2}.\eqno(2.27)$$
As a consequence the energy-momentum tensor (2.23) has off-diagonal elements $T^{01}$. When we consider the late universe where
$a(t)$ is big, one may neglect $u^1$ (2.27) and we are back at standard cosmology. In the following we investigate the first correction to standard cosmology assuming the energy-momentum tensor for a matter dominated universe with $p=0$:
$$T^{00}=\ro (u^0)^2=\ro(1+{1\over K^2a^2})\eqno(2.28)$$
$$T^{01}=\ro u^0u^1={\ro\over Ka^2}\sqrt{1+{1\over K^2a^2}}\eqno(2.29)$$
$$T^{11}=\ro(u^1)^2={\ro\over K^2a^4}\eqno(2.30)$$
and zero otherwise.

\section{Modification of the Robertson-Walker metric} 

Since the energy-momentum tensor (2.32) has off-diagonal components so must have the metric tensor $g_{\mu\nu}$ as well. Therefore we modify the Robertson-Walker metric as follows
$$ds^2=dt^2+2b(t)dt\,dr-a^2(t)[dr^2+r^2(d\te^2+\sin^2\te d\phi^2)].\eqno(3.1)$$
We will find that $b(t)$ and $a(t)$ still depend only on time, when we consider the first nonstandard correction. The components of the inverse metric are equal to
$$g^{00}={a^2\over D},\quad g^{01}={b\over D},\quad g^{11}=-{1\over D}\eqno(3.2)$$
$$g^{22}=-{1\over a^2r^2},\quad g^{33}=-{1\over a^2r^2\sin^2\te}\eqno(3.3)$$
where
$$D=a^2+b^2\eqno(3.4)$$
is the determinant of the $2\times 2$ matrix of the $t$, $r$ components. The non-vanishing Christoffel symbols are now given by
$$\Gamma^0_{00}={b\dot b\over D},\quad \Gamma^0_{01}=-{ab\over D}\dot a,\quad
\Gamma^0_{11}={a^3\over D}\dot a$$
$$\Gamma^0_{22}={a^3\over D}\dot ar^2+{b\over D}a^2r,\quad\Gamma^0_{33}=\B({a^3\over D}\dot a r^2+{b\over D}a^2r\B)\sin^2\te
\eqno(3.5)$$
$$\Gamma^1_{00}=-{\dot b\over D},\quad\Gamma^1_{01}={a\dot a\over D},\quad \Gamma^1_{11}={ab\over D}\dot a\eqno(3.6)$$
$$\Gamma^1_{22}=-{a^2\over D}r+r^2b{a\dot a\over D},\quad\Gamma^1_{33}=\B(-{a^2\over D}r+r^2b{a\dot a\over D}\B)\sin^2\te,\eqno(3.7)$$
$$\Gamma^2_{02}={\dot a\over a},\quad\Gamma^2_{12}={1\over r},\quad\Gamma^2_{33}=-\sin\te\cos\te\eqno(3.8)$$
$$\Gamma^3_{03}={\dot a\over a},\quad\Gamma^3_{13}={1\over r},\quad\Gamma^3_{23}={\cos\te\over\sin\te}.\eqno(3.9)$$

Next we calculate the Ricci tensor
$$R_{\mu\nu}=\d_\al\Gamma^\al_{\mu\nu}-\d_\nu\Gamma^\al_{\mu\al}+\Gamma^\beta_{\mu\nu}\Gamma^\al_{\al\beta}
-\Gamma^\al_{\nu\beta}\Gamma^\beta_{\al\mu}.$$
We obtain
$$R_{00}=-{\ddot a\over D}\B(3a+2{b^2\over a}\B)-{b^2\dot a^2\over D^2}+{\dot a\dot b\over D^2}\B(3ab+2{b^3\over a}\B)
-2{\dot b\over rD}\eqno(3.10)$$
$$R_{01}=-\ddot a{ab\over D}-\dot a^2{b\over D^2}(2a^2+3b^2)+{ab^2\over D^2}\dot a\dot b-{2b^2\dot a\over raD}.\eqno(3.11)$$
$$R_{11}={a^3\over D}\ddot a+{a^4\dot a^2\over D^2}\B(2+3{b^2\over a^2}\B)-{a^3b\over D^2}\dot a\dot b
+{2\over r}{a\dot a\over D}b\eqno(3.12)$$
$$R_{22}=r^2\B[{a^3\over D}\ddot a+{a^4\dot a^2\over D^2}\B(2+3{b^2\over a^2}\B)-{a^3b\over D^2}\dot a\dot b\B]
+{b^2\over D}+r{ab\over D^2}\dot a(3a^2+4b^2)+r{a^4\over D^2}\dot b\eqno(3.13)$$
$$R_{33}=R_{22}\sin^2\te.\eqno(3.14)$$
This gives the following scalar curvature
$$R=g^{\mu\nu}R_{\mu\nu}=-6{a\ddot a\over D}+{\dot a^2\over D^2}(-6a^2-12b^2)+$$
$$+6\dot a\dot b{ab\over D^2}-{4\over rD^2}\B(a^2\dot b+2ab\dot a+3{b^3\over a}\dot a\B)
-2{b^2\over r^2a^2D}.\eqno(3.15)$$

Now we are ready to calculate the Einstein tensor:
$$G_{00}=R_{00}-{g_{00}\over 2}R={\dot a\over D^2}\B(3a^2+5b^2\B)-2{\ddot ab^2\over aD}+$$
$$+2{\dot a\dot b\over D^2}{b^3\over a}-2{\dot b\over rD^2}b^2+{\dot ab\over rD^2}(2a+3{b^2\over a})
+{b^2\over r^2a^2D}.\eqno(3.16)$$
We shall soon see that $b=O(1)$. Then the first term on the righthand side is $O(1)$ and the corrections are $O(a^{-2})$.
The off-diagonal component is equal to
$$G_{01}=R_{01}-{g_{01}\over 2}R=b\B(2{\ddot a\over D}+{a\over D23}\dot a^2\B)a.\eqno(3.17)$$
Here we have only written down the terms $O(1)$. They give the first nonstandard correction. 
The remaining components of the Einstein tensor are equal to
$$G_{11}=-2{\ddot aa^3\over D}+{a^2\dot a^2\over D^2}\B(-a^2-3b^2\B)
+2\dot a\dot b{a^3b\over D^2}+$$
$$-{\dot a\over rD^2}(2a^3b+4ab^3)-{2a^4\over rD^2}\dot b-{b^2\over r^2D}.\eqno(3.18)$$
$$G_{22}=r^2\B[-2{a^3\over D}\ddot a-{a^2\dot a^2\over D^2}\B(a^2+3b^2\B)+$$
$$+2{a^3b\over D^2}\dot a\dot b-{a\over rD^2}(a^2b\dot a+a^3\dot b+2b^3\dot a)\B].\eqno(3.19)$$

For the Einstein's equations we now need $T_{\mu\nu}$ with lower indices. From equations (2.28-30) we obtain
$$T_{00}=(g_{00})^2T^{00}+2g_{00}g_{01}T^{01}+(g_{01})^2T^{11}=\ro(1+O(a^{-2})\eqno(3.20)$$
$$T_{01}=g_{00}g_{10}T^{00}+(g_{00}g_{11}+g_{01}^2)T^{01}+g_{01}g_{11}T^{11}=$$
$$=\ro\B[b(1+{1\over K^2a^2})+{b^2-a^2\over Ka^2}\sqrt{1+{1\over K^2a^2}}-{a^2b\over K^2a^4}\B]=\ro[b-{1\over K}+O(a^{-2})]
\eqno(3.21)$$
$$T_{11}=(g_{10})^2T^{00}+2g_{00}g_{10}T^{01}+(g_{11})^2T^{11}=\ro\B(b^2-{2b\over K}+O(a^{-2}\B).\eqno(3.22)$$
We first solve the equation
$$G_{11}=8\pi GT_{11}.$$
Since $T_{11}$ (3.22) is $O(1)$ we have in leading order $O(a^2)$
$$G_{11}=-2{a^5\over a^4}\ddot a-\dot a^2=0.$$
This agrees with standard cosmology and gives the usual scale function for the matter-dominated universe
$$a(t)=a_1 t^{2/3},\eqno(3.23)$$
where $a_1$ is an integration constant. Next we determine the density $\ro$ from
$$G_{00}=3{\dot a^2\over a^2}+O(a^{-2})=8\pi G\ro.$$
With the result (3.23) we obtain
$$\ro={1\over 6\pi Gt^2}.\eqno(3.24)$$
Now we use these results in
$$G_{01}=8\pi GT_{01}.\eqno(3.25)$$
From equations (3.17) and (3.21) we obtain
$${b\over a}(2\ddot a+{\dot a^2\over a})=8\pi G\ro(b-{1\over K})\eqno(3.26)$$
where we have again neglected $O(a^{-2})$. Using $a(t)$ (3.23) we see that the lefthand side vanishes, therefore we must have
$$b(t)={1\over K}.\eqno(3.27)$$
The remaining Einstein's equations are satisfied to leading order in virtue of equation (3.23). The corrections are of order $O(a^{-2})$ and are not considered here, as far as the metric is concerned.

The modified Robertson-Walker line element is now equal to
$$ds^2=dt^2+{1\over K}dt\,dr-a^2(t)[dr^2+r^2(d\te^2+\sin^2\te d\phi^2)].\eqno(3.28)$$
The new constant $K$  appears in the modified metric which again shows its universality.
It is interesting to consider the new radial geodesics. Instead of equation (2.19) we now have
$${dv_r\over dt}+2{a\dot a\over D}v_r+3{ab\over D}\dot av_r^2-{a^3\dot a\over D}v_r^3=0\eqno(3.29)$$
because $\dot b=0$ by (3.27). On the other hand the radial null curves for the metric (3.28) are given by
$$c_1={1\over a^2}(b\pm\sqrt{D}).\eqno(3.30)$$
Using $\dot b=0$ it is straight-forward to check that this is a special solution of equation (3.29) which
again describes light rays in next to leading order. The complete integration of this equation (3.29) is not simple.

\section{Maxwell's equations, redshift and luminosities}

Maxwell's equations in a gravitational field are obtained by replacing the Minkowski tensor by $g_{\mu\nu}$ and all derivatives by covariant derivatives [4]. Then Maxwell's equations without current-sources read
$$\nabla_\mu F^{\mu\nu}=0\eqno(4.1)$$
$$\nabla_\lambda F_{\mu\nu}+\nabla_\nu F_{\lambda\mu}+\nabla_\mu F_{\nu\lambda}=0.\eqno(4.2)$$
Since we work in spherical coordinates the electromagnetic field tensor $F^{\mu\nu}$ must be expressed in these coordinates as well.
The field tensor in Cartesian coordinates has the well known form
$$\tilde F^{\mu\nu}=\pmatrix{0&-E^x&-E^y&-E^z\cr E^x&0&-B^z&B^y\cr E^y&B^z&0&-B^x\cr E^z&-B^y&B^x&0\cr}.\eqno(4.3)$$
Our convention of the metric etc. agrees with [6]. We transform to spherical coordinates
$$x=r\sin\te\cos\phi,\quad y=r\sin\te\sin\phi,\quad z=r\cos\te$$
using the inverse transformation
$$r=\sqrt{x^2+y^2+z^2},\quad \te=\arctan{{\sqrt{x^2+y^2}\over z}},\quad \phi=\arctan{{y\over x}}.\eqno(4.4)$$
The transformed field tensor is given by
$$F^{\mu\nu}(x)=\tilde F^{\al\beta}(\tilde x){\d x^\mu\over\d\tilde x^\al}{\d x^\nu\over\d\tilde x^\beta},\eqno(4.5)$$
where $\tilde x^\al$ are the Cartesian coordinates. The result is
$$F^{\mu\nu}=\pmatrix{0&-E^r&-E^\te&-E^\phi\cr E^r&0&-B^\phi\sin\te&B^\te/\sin\te\cr E^\te&B^\phi\sin\te&0&-B^r/(r^2\sin\te)\cr E^\phi&-B^\te/\sin\te&B^r/(r^2\sin\te)&0\cr},\eqno(4.6)$$
where
$$E^r=E^x\sin\te\cos\phi+E^y\sin\te\sin\phi+E^z\cos\te$$
$$E^\te={1\over r}(E^x\cos\te\cos\phi+E^y\cos\te\sin\phi-E^z\sin\te)\eqno(4.7)$$
$$E^\phi={1\over r\sin\te}(-E^x\sin\phi+E^y\cos\phi)$$
and the same relations for the magnetic components. The electric components (4.7) are what one would expect for a true 3-vector. But the magnetic components in (4.6) have additional factors because the magnetic field $\vec B$ is a polar vector.

Now the Maxwell's equations (4.1) read
$$\nabla_\mu F^{\mu 0}=\d_\mu F^{\mu 0}+\Gamma^\mu_{\mu\lambda}F^{\lambda 0}=0=$$
$$=\d_rE^r+\d_\te E^\te+\d_\phi E^\phi+{2\over r}E^r+{\cos\te\over\sin\te}E^\te\eqno(4.8)$$
$$\nabla_\mu F^{\mu 1}=\d_\mu F^{\mu 1}+\Gamma^\mu_{\mu\lambda}F^{\lambda 1}=0=$$
$$=-\d_tE^r+\sin\te\d_\te B^\phi-{1\over\sin\te}\d_\phi B^\te-2{\dot a\over a}E^r-{a\dot a\over D}E^r\eqno(4.9)$$
$$\nabla_\mu F^{\mu 2}=\d_\mu F^{\mu 2}+\Gamma^\mu_{\mu\lambda}F^{\lambda 2}=0=$$
$$=-\d_tE^\te-\sin\te \d_r B^\phi+{1\over r^2\sin\te}\d_\phi B^r-\B({a\dot a\over D}+2{\dot a\over a}\B)E^\te-2{\sin\te\over r}B^\phi\eqno(4.10)$$
$$\nabla_\mu F^{\mu 3}=\d_\mu F^{\mu 3}+\Gamma^\mu_{\mu\lambda}F^{\lambda 3}=0=$$
$$=-\d_tE^\phi+{1\over\sin\te}\d_r B^\te-{1\over r^2}\d_\te {B^r\over\sin\te}-\B({a\dot a\over D}+2{\dot a\over a}\B) E^\phi+(\Gamma^2_{21}+\Gamma^3_{31}){B^\te\over\sin\te}-$$
$$-\Gamma^3_{32}{B^r\over r^2\sin\te}.\eqno(4.11)$$
Here we have used the fact that
$$\Gamma^0_{01}+\Gamma^1_{11}=0.$$

For the homogeneous equations (4.2) we need the components with lower indices
$$F_{10}=-DE^r,\quad F_{20}=-a^2r^2(E^\te+bB^\phi\sin\te)\eqno(4.12)$$
$$F_{30}=-a^2r^2\sin^2\te(E^\phi-b{B^\te\over \sin\te})\eqno(4.13)$$
$$F_{12}=-a^2r^2(-bE^\te+a^2B^\phi\sin\te)\eqno(4.14)$$
$$F_{13}=-a^2r^2\sin^2\te(-bE^\phi-a^2{B^\te\over \sin\te}),\quad F_{23}=-a^4r^2\sin\te B^r.\eqno(4.15)$$
It is well known that in the homogeneous equations (4.2) the covariant derivative can be replaced by ordinary partial derivatives [6]. 
In the following we consider one simple multipole radiation with $B^r=0=B^\te=E^\phi$; the general case will be investigated in a later paper. The radiation is emitted by a distant star.
So we choose the origin $r=0$ of our spherical coordinates in the center of the star, and for the moment we neglect its motion.
Then the $(0,1,2)$-component of (4.2) becomes
$$\sin\te\d_tB^\phi={b\over a^2}\d_0E^\te-{1\over a^2}(\d_rE^\te+b\sin\te\d_rB^\phi)+2{b\dot a\over a^3}E^\te-$$
$$-{4\dot a\over a}B^\phi\sin\te-{2\over a^2r}(E^\te+bB^\phi\sin\te)-{1\over a^2r^2}\B(1+{b^2\over a^2}\B)\d_\te E^r=0.\eqno(4.16)$$
The other homogeneous equations are trivially satisfied in our case, so that we must solve (4.9), (4.10) and (4.16). 

We assume the following simple angular dependence:
$$E^\te(t,r,\te)=\tilde E^\te(t,r)\sin\te,\quad E^r(t,r,\te)=\tilde E^r(t,r)\cos\te\eqno(4.17)$$
and $B^\phi=B^\phi(t,r)$ is independent of $\te$. Now equation (4.10) gives
$$\d_t\tilde E^\te+\d_rB^\phi+\B({a\dot a\over D}+2{\dot a\over a}\B)\tilde E^\te+{2\over r}B^\phi=0.\eqno(4.18)$$
This equation must be combined with equation (4.16)
$$\d_tB^\phi={b\over a^2}\d_t\tilde E^\te-{1\over a^2}(\d_r\tilde E^\te+b\d_rB^\phi)-\B(4{\dot a\over a}+{2b\over a^2r}\B)B^\phi+$$
$$+\B(2{\dot ab\over a^3}-{2\over a^2r}\B)\tilde E^\te+{1\over a^2r^2}\B(1+{b^2\over a^2}\B)\d_\te\tilde E^r.\eqno(4.19)$$
We further simplify the equations by the substitutions
$$\tilde E^r={1\over r}f_1(t,r),\quad\tilde E^\te={1\over r^2}f(t,r),\quad B^\phi={1\over r^2}g(t,r).\eqno(4.20)$$
Then we have
$$\d_tg={b\over a^2}\d_tf-{1\over a^2}(\d_rf+b\d_rg)-4{\dot a\over a}g+2{\dot ab\over a^3}f+{1\over a^2r}\B(1+{b^2\over a^2}\B)
f_1.\eqno(4.21)$$
and equation (4.18) becomes
$$\d_tf+\d_rg+3{\dot a\over a}f=0.\eqno(4.22)$$
Here we have neglected quadratic corrections so that we put $D=a^2+O(1)$. From (4.9) we get
$$\d_tf_1-2{g\over r}+3{\dot a\over a}f_1=0.\eqno(4.23)$$
A last simplification is obtained by the substitutions
$$f_1={\tilde f_1\over a^3},\quad f={\tilde f\over a^3},\quad g={\tilde g\over a^4}.\eqno(4.24)$$
Now the following three equations remain to be solved
$$\d_t\tilde f+{1\over a}\d_r\tilde g=0$$
$$\d_t\tilde g+{1\over a}\d_r\tilde f={b\over a}\d_t\tilde f-{b\over a^2}\d_r\tilde g-{\dot ab\over a^2}\tilde f
+{1\over ar}\B(1+{b^2\over a^2}\B)\tilde f_1..\eqno(4.25)$$
$$\d_t\tilde f_1=2{\tilde g\over ar}.$$

We solve the above equations in the eikonal approximation by using the following ansatz
$$\tilde f=f_0(t)\sin[k\fii(t,r)],\quad \tilde g=g_0(t)\sin[k\fii(t,r)]\eqno(4.26)$$
$$\tilde f_1=f_2(t)\sin[k\fii(t,r)]$$
In this approximation $k$ is large, that means the wavelength of the radiation is small compared to cosmic scales which is extremely well satisfied in cosmology. Then in all equations we take only the contributions $O(k)$ which come from differentiating $\sin(k\fii)$. We then obtain from equation (4.25)
$$f_0\d_t\fii+{g_0\over a}\d_r\fii=0,\eqno(4.27)$$
$$\B(g_0-{b\over a}f_0\B)\d_t\fii+\B({f_0\over a}+{b\over a^2}g_0\B)\d_r\fii=0.$$
The last equation (4.25) leads to
$$f_2\d_t\fii=0.\eqno(4.28)$$
This implies $f_2=0$ i.e. $E^r$ vanishes in this approximation.

For a nontrivial solution the determinant in (4.27) must vanish:
$$f_0\B({f_0\over a}+{bg_0\over a^2}\B)-{g_0\over a}\B(g_0-{bf_0\over a}\B)=0.$$
This gives the relation
$${g_0\over af_0}=\pm\sqrt{{a^2+b^2\over a^4}}+{b\over a^2}\equiv c_2(t)\eqno(4.29)$$
and the single ''eikonal`` equation
$${1\over c_2}\d_t\fii+\d_r\fii=0.\eqno(4.30)$$
This shows that $c_2(t)$ (4.29) is the phase velocity of the radiation which depends on the gravitational field through $a(t)$. In leading order $c_2$ agrees with the signal or group velocity $c_1$ (3.31). It is known that the gravitational field has some similarity with a dispersive optical medium [6]. The plus sign in equation (4.29) corresponds to outgoing radiation which we consider in the following. We note the boundary condition
$$c_2(T)=c\eqno(4.31)$$
where $T$ is the age of the universe and $c$ the present light speed measured on earth. Now the general solution of equation (4.30) is
$$\fii(t,r)=F\B(\int\limits_T^tc_2(t')dt'+cT-r\B)\eqno(4.32)$$
where $F$ is some differentiable function. For monochromatic light we put
$$k\fii=\omega t-kr=k(\int\limits_T^tc_2(t')dt'+cT-r)\eqno(4.33)$$
where
$$\omega(t)={k\over t}\B(\int\limits_T^t c_2(t')dt'+cT\B).\eqno(4.34)$$
On earth this gives the usual relation 
$$\omega(T)=ck.\eqno(4.35)$$
The comparison of frequencies at different cosmic times is given by
$${\omega(t)\over\omega(T)}={1\over ct}\B(\int\limits_T^t c_2(t')dt'+cT\B).\eqno(4.36)$$

Until now we have calculated in the rest frame of the emitting star which in reality is in radial motion away from us. We must transform the results to the cosmic rest frame with coordinates $x'=(t',r',\te,\phi)$ by the transformation
$$t'=t,\quad r'=r-\int\limits_T^tv_r(\tau)d\tau,\quad\te'=\te,\quad\phi'=\phi.\eqno(4.37)$$
To understand the minus sign, note that $r'>r$ for $t<T$. The field tensor is transformed according to 
$$F'^{\mu\nu}(x')={\d x'^\mu\over\d x^\al}{\d x'^\nu\over\d x^\beta}F^{\al\beta}(x).\eqno(4.38)$$
Instead of equation (4.33) we then have
$$k\fii=\omega t-kr=k\B(\int\limits_T^t(c_2-v_r)(\tau)d\tau+cT-r'\B).\eqno(4.39)$$
Now instead of the relation (4.36) we obtain the ratio between the emitted and observed frequencies
$${\omega_{\rm em}\over\omega_{\rm obs}}={1\over ct}\B(\int\limits_T^t(c_2-v_r)(\tau)d\tau+cT\B)\equiv 1+z.\eqno(4.40)$$
This gives the redshift $z$. To compare the result (4.40) with the standard expression we expand the integrand in powers of $1/a$
$$c_2-v_r={1\over a}-{3\over Ka^2}+\ldots$$
The first term agrees with standard cosmology, it comes from the action of the gravitational field alone on the radiation. The next term $O(a^{-2})$ contains the Doppler effect due to the motion of the emitting star.

To calculate the absolute and apparent luminosities we must compute the Poynting vector. We start from the electromagnetic energy-momentum tensor
$$T^{\mu\nu}_\el=-F^\mu_{\,\lambda}F^{\nu\lambda}+{1\over 4}g^{\mu\nu}F_{\al\beta}F^{\al\beta}.$$
Again we first calculate in the rest frame of the star. We find
$$F_{\al\beta}F^{\al\beta}=a^2r^2[(E^\te)^2+a^2(B^\phi)^2\sin^2\te-2bE^\te B^\phi\sin\te]$$
and using
$$E^\te=f_0{\sin\te\over a^3r^2}\sin(\omega t-kr)$$
$$B^\phi=f_0{c_2\over a^3r^2}\sin(\omega t-kr)\eqno(4.41)$$
we get
$$T^{00}_\el=-g_{22}(F^{02})^2+{1\over 4}g^{00}F_{\al\beta}F^{\al\beta}=$$
$$={f_0^2\over a^4r^2}\B(1+{a^2\over 2D}+c_2^2{a^4\over 2D}-c_2{ba^2\over D}\B)\sin^2\te\sin^2(\omega t-kr).
\eqno(4.42)$$
The off-diagonal component is
$$T^{01}_\el=-g_{22}F^{02}F^{12}+{1\over 4}g^{01}F_{\al\beta}F^{\al\beta}=$$
$$={f_0^2\over a^4r^2}\B[c_2+{b\over 2D}(1+a^2c_2^2)\B]\sin^2\te\sin^2(\omega t-kr)
\eqno(4.43)$$
The factor $\sin^2\te/r^2$ shows that we consider dipole radiation, the dipole oscillates in the $z$-direction. Of course, in a star which collapses in a supernova, giving a good standard candle for distance measurements, there are dipoles in all directions. Therefore, we average $\sin^2\te$ and also $\sin^2(\omega t-kr)$, both giving a factor 1/2 in the final result. As in (4.38) we must transform the result to the cosmic rest frame
$$T'^{01}_\el={\d x'^0\over\d x^\al}{\d x'^1\over\d x^\beta}T^{\al\beta}_\el=v_rT^{00}_\el+T^{01}_\el=$$
$$={f_0^2\sin^2\te\over a^4r^2}\sin^2(\omega t-kr)\B\{{1\over a}(1+{b\over a})+{b\over D}+v_r\B[1+{a^2\over D}(1+{b\over a})
  -{ba\over D}\B]\B\}.\eqno(4.44)$$
The leading order comes from $T^{01}_\el$.

We want to show that the tensor (4.44) gives the desired luminosities without any additional scale factor $a(t)$. The electromagnetic
energy-momentum tensor is conserved
$$\nabla_\mu T^{\mu\nu}_\el=\d_\mu T^{\mu\nu}_\el+\Gamma^\mu_{\mu\lambda}T^{\lambda\nu}_\el+\Gamma^\nu_{\mu\lambda}
T^{\mu\lambda}_\el=0.\eqno(4.45)$$
For $\nu=0$ we have energy conservation
$$\d_tT^{00}_\el=-\d_jT^{j0}_\el-\Gamma^\mu_{\mu\lambda}T^{\lambda 0}_\el-\Gamma^0_{\mu\lambda}
T^{\mu\lambda}_\el.$$
Now we integrate over a spatial volume $V$ containing the star and apply the 3-dimensional Gauss' theorem in the first term on the right
$$\d_t\int\limits_VT^{00}_\el d^3x=-\int\limits_{\d V} T^{j0}_\el d\sigma_j-\int\limits_V(\Gamma^\mu_{\mu\lambda}T^{\lambda 0}_\el+\Gamma^0_{\mu\lambda}T^{\mu\lambda}_\el)d^3x.\eqno(4.46)$$
The surface integral on the right is the electromagnetic energy flux which flows to infinity, this gives the absolute luminosity of the star. The last volume integral in (4.46) must be interpreted as an interaction energy between the electromagnetic and gravitational fields [4].

After averaging over the angle $\te$ and time, the surface integral is trivial if we consider a sphere and gives a factor $4\pi r^2$. Then the absolute luminosity $L$ is equal to
$$L={\pi f_0^2\over a^5}\B(1+ 2{b\over a}+2av_r\B)\eqno(4.47)$$
up to higher order terms. Here the time $t$ of emission of the radiation appears in $a(t)$ and $v_r(t)$. The apparent luminosity $l$ is equal to the result (4.44) averaged, but taken at present time $T$. This gives for the luminosity distance $d_L$
$$d_L^2={L\over 4\pi l}=r^2{a^5(T)\over a^5(t)}{1+(2{b\over a}+2av_r)(t)\over 1+(2{b\over a}+2av_r)(T)}$$
or
$$d_L(t)=r(t)\biggl[{a^5(T)\over a^5(t)}{1+(2{b\over a}+2av_r)(t)\over 1+(2{b\over a}+2av_r)(T)}\biggl]^{1/2}.\eqno(4.48)$$
This differs considerably from standard cosmology [5].
The coordinate distance $r(t)$ has to be calculated by integrating equation (2.22)
$$r(t)=\int\limits_t^T v_r(t')dt'.\eqno(4.49)$$

\section{Connection with observations}

One most interesting observable is the age $T$ of the universe or the Hubble constant $H_0$. The model-independent definition of $H_0$ is given by the expansion
$$d_L(z)={c\over H_0}\B[z+{1\over 2}(1-q_0)z^2+\ldots\B],\eqno(5.1)$$
where $q_0$ is the deceleration parameter. We emphasize that in nonstandard cosmology the much used definition [5]
$$\tilde H_0={\dot a\over a}\Bigl\vert_T\eqno(5.2)$$
would give a different quantity which is not directly related to $z$ and $d_L$. We calculate $H_0$ from
$${c\over H_0}={d\,d_L\over dz}\Bigl\vert_{z=0}={d\,d_L\over dt}\Bigl\vert_T\Bigl({dz\over dt}\B)^{-1}\Bigl\vert_T\eqno(5.3)$$
using our results (4.40) and (4.48). Since
$${dz\over dt}=-{1\over ct^2}\B(\int\limits_T^t(c_2-v_r)d\tau+cT\B)+{1\over ct}(c_2-v_r)(t)\eqno(5.4)$$
we obtain
$${dz\over dt}\Bigl\vert_T=-{v_r\over cT}=-{c\over K T}=-{1\over a_1 K}T^{-5/3}.\eqno(5.5)$$
Here we have calculated in leading order using equations (4.31) (4.29) and (2.22).

For the luminosity distance we need
$$r(t)=\int\limits_t^Tv_r(\tau)d\tau={3\over a_1^2 K}(t^{-1/3}-T^{-1/3})\eqno(5.6)$$
where again we have computed in leading order $v_r\sim a^{-2}$.
Then from equation (4.48) we get
$$d_L(t)={3\over a_1^2 K}T^{5/3}\B({1\over t^2}-{1\over T^{1/3}t^{5/3}}\B)\eqno(5.7)$$
and
$${dd_L\over dt}={3\over a_1^2 K}T^{5/3}\B(-{2\over t^3}+{5\over 3}{T^{-1/3}\over t^{8/3}}\B).\eqno(5.8)$$
For $t=T$ we find
$${d\,d_L\over dt}\B\vert_T=-{1\over a_1^2 KT^{4/3}}.\eqno(5.9)$$
Substituting into equation (5.3) and using the results (4.31) and (3.23) we arrive at
$${c\over H_0}={T^{1/3}\over a_1}={T\over a_1T^{2/3}}=cT$$
or
$$T={1\over H_0}.\eqno(5.10)$$
With the standard value $H_0=70$ km/(s Mpc) this gives the correct age 14 Gyr of the universe. In standard cosmology one needs a cosmological constant $\Lambda$ to achieve this [5].

The new constant $K$ has cancelled in the Hubble constant (5.3). This is not the case when we calculate the deceleration parameter $q_0$. According to the definition (5.1) we must compute
$${d^2d_L\over dz^2}={d\over dt}\B({dd_L\over dz}\B)\B({dz\over dt}\B)^{-1}={d\over dt}\B[{dd_L\over dt}\B({dz\over dt}\B)^{-1}\B]\B({dz\over dt}\B)^{-1}=$$
$$={d^2d_L\over dt^2}\B({dz\over dt}\B)^{-2}-{dd_L\over dt}\B({dz\over dt}\B)^{-3}{d^2z\over dt^2}.\eqno(5.11)$$
From equation (5.4) we obtain
$${d^2z\over dt^2}={2\over ct^3}\B(\int\limits_T^t(c_2-v_r)d\tau+cT\B)-{2\over ct^2}(c_2-v_r)+{1\over cT}(\dot c_2-\dot v_r).$$
Here we again substitute the time dependence (2.22) in leading order. Then for $t=T$ we finally arrive at
$${d^2z\over dt^2}\B\vert_T={1\over T^2}\B(-{2\over 3}+{10\over 3}{c\over K}\B).\eqno(5.12)$$
The constant $a_1$ is always eliminated by means of equations (4.31) and (3.23)
$$a_1={T^{-2/3}\over c}.\eqno(5.13)$$

Next we differentiate equation (5.8) and set $t=T$ giving
$${d^2d_L\over dt^2}\B\vert_T={14\over 3a_1^2 K}T^{-7/3}.\eqno(5.14)$$
Now we are ready to substitute everything into equation (5.11) for $t=T$ or $z=0$. Then using the definition (5.1) we arrive at
$${d^2d_L\over dz^2}\B\vert_{z=0}={c\over H_0}(1-q_0)=$$
$$={4\over 3} KT+{2\over 3c}K^2T.\eqno(5.15)$$
Finally the deceleration parameter is given by
$$1-q_0={2\over 3c}\B(2 K+{K^2\over c}\B).\eqno(5.16)$$
Consequently, the new constant $K$ has a direct observational meaning. For the special value
$$ K=c\B(\sqrt{{5\over 2}}-1\B)\equiv K_0\eqno(5.17)$$
we get $q_0=0$. For $K>K_0$ one has $q_0<0$ so that the expansion is accelerating as indicated by the supernova data. However, taking this as evidence for dark energy seems to be jumping to the conclusion.

So far so good. However, nonstandard cosmology also has a missing mass problem. It can be seen from the lowest order result (3.24) for the matter density $\ro$
$$\ro={1\over 6\pi Gt^2}.\eqno(5.18)$$
In virtue of equation (3.10) we then have for $t=T$
$$\ro(T)={H_0^2\over 6\pi G}.\eqno(5.19)$$
In standard cosmology one defines a critical density [5]
$$\ro_{\rm crit}={3H_0^2\over 8\pi G}= 0.409\times 10^{-30} {\rm g/cm^3}\eqno(5.20)$$
for $H_0=70$ km/(s Mpc). That means the present matter density (5.18) is 4/9 of the critical density, which is considerably more than the observed density of ordinary baryonic matter. However, the result (5.18) is only true in lowest order. A glance at the equations shows that in $O(a^{-2})$ one has to get rid of the $r$-dependence. We see that the early universe is much more complicated in nonstandard cosmology than it is in the standard theory, because it is not homogeneous. This interesting problem will be investigated in a later paper.

\end{document}